\documentclass[11pt]{article}
\begin{document}
\setcounter{page}{1}
\def\theequation{\arabic{section}.\arabic{equation}}
\def\theequation{\thesection.\arabic{equation}}
\setcounter{section}{0}

\title{On the $K^{\pm}$--meson production\\ from the
quark--gluon plasma phase in ultra--relativistic heavy--ion
collisions}

\author{A. Ya. Berdnikov, Ya. A. Berdnikov~\thanks{E--mail:
berdnikov@twonet.stu.neva.ru, State Technical University, Department
of Nuclear Physics, 195251 St. Petersburg, Russian Federation} ,\\
A. N. Ivanov~\thanks{E--mail: ivanov@kph.tuwien.ac.at, Tel.:
+43--1--58801--14261, Fax: +43--1--58801--14299}~$^\P$ ,
V. F. Kosmach~\thanks{State Technical University, Department of
Nuclear Physics, 195251 St. Petersburg, Russian Federation} ,
V. M. Samsonov~\thanks{E--mail: samsonov@hep486.pnpi.spb.ru,
St.Petersburg Institute for Nuclear Research, Gatchina, Russian
Federation} , N. I. Troitskaya~\thanks{Permanent Address: State
Technical University, Department of Nuclear Physics, 195251
St. Petersburg, Russian Federation}}

\date{\today}

\maketitle

\begin{center}
{\it Institut f\"ur Kernphysik, Technische Universit\"at Wien, \\
Wiedner Hauptstr. 8-10, A-1040 Vienna, Austria}
\end{center}

\begin{center}
\begin{abstract}
An abundance of the strangeness that can be induced in a thermalized
quark--gluon plasma (QGP) is considered as a signal of the QGP phase
appearing in the intermediate state of ultra--relativistic heavy--ion
collisions. As a quantitative characteristic of this signal we take
the ratio $R_{K^+ K^-} = N_{K^+}/N_{K^-}$ of the multiplicities of the
production of $K^{\pm}$ mesons. This ratio is evaluated for a
thermalized QGP phase of QCD and for the quark--gluon system escaped
from the QGP phase.  For a thermalized QGP phase the ratio $R_{K^+
K^-}$ has been found as a smooth function of a 3--momentum of the
$K^{\pm}$ mesons and a temperature ranging the values from the region
$160\,{\rm MeV} < T <200\,{\rm MeV}$. We show that at the temperature
$T= 175\, {\rm MeV}$ our prediction for the ratio $R_{K^+K^-}(q,T =
175) = 1.80^{+ 0.04}_{-0.18}$ agrees good with the experimental data
of NA49 and NA44 Collaborations on central ultra--relativistic Pb--Pb
collisions at 158 GeV per nucleon: $R^{\exp}_{K^+K^-} = 1.80\pm
0.10$. For the ratio of the $K^+$ and $\pi^+$ multiplicities we have
obtained the value $R_{K^+\pi^+}(q,T = 175) = 0.134\pm 0.014$ agreeing
good with the experimental data of NA35 Collaboration on the
nucleus--nucleus collisions at 200\,GeV per nucleon
$R^{\exp}_{K^+\pi^+} = 0.137\pm 0.008$.\\
\vspace{0.2in}
\end{abstract}
\end{center}

\newpage

\section{Introduction}
\setcounter{equation}{0}

Nowadays there is a consensus that QCD gives a satisfactory description of
strong interactions of hadrons. The important question still left concerns
the properties of the QCD vacuum. One of the approaches to the exploration
of the properties of the QCD vacuum is to investigating  the excited
vacuum states at high densitities and temperature. The quark--gluon plasma
(QGP) phase of QCD [1,2] is just the excited QCD vacuum in which quarks,
anti--quarks and gluons being at the deconfined phase collide frequently
each other. There is a belief [2] that the QGP phase of the quark--gluon
system can be realized in ultra--relativistic heavy--ion collision
($E_{\rm cms}/{\rm nucleon} \gg 1\,{\rm GeV}$) experiments.

There are two main approaches to the description of the QGP phase --
the dynamical [3] and the thermal [4] ones. The dynamical approach
employs the methods related to the relativistic kinetic equations
supplemented by semi--hard interactions at very high energies. The
thermal approach is based on the ideal gas assumption supposing a
thermalization of the QGP after some initial time $\tau_0$. Finally,
in both approaches the QGP phase evolutes to the hadronic phase.

The difficulties arising from the recognition of the QGP, induced in
ultra--relativistic heavy--ion collisions, through a hadronization
concern the following. The hadrons can also be produced in heavy--ion
collisions by the quark--gluon system escaped from the QGP
phase. Therefore, one needs to have distinct criteria allowing to
distinguish the hadrons produced by the QGP phase from the hadrons
procreated by the quark--gluon system escaped from the intermediate
QGP phase.

As has been suggested in Refs.[5,6] in a thermalized QGP one can
expect an abundance of the production of strange hadrons $K$,
$\Lambda$, $\Xi$ and so on. Such an abundance can serve as a criterion
for the formation of the QGP [5,6]. The arguments for the enhancement
of the strange hadron production are the following. At very high
energies of heavy--ion collisions the quark--gluon system is composed
from highly relativistic and very dense quarks, anti--quarks and
gluons. By virtue of the asymptotic freedom the particles are almost
at liberty and due to high density collide themselves frequently that
leads to an equilibrium state. If to consider such a state as a
thermalized QGP phase of QCD, the probabilities of light
massless quarks $n_q(\vec{p}\,)$ and light massless anti--quarks
$n_{\bar{q}}(\vec{p}\,)$, where $q = u$ or $d$, to have a momentum $p$
at a temperature $T$, can be described by the Fermi--Dirac
distribution functions [1,4,7]:
\begin{eqnarray}\label{label1.1}
n_q(\vec{p}\,) = \frac{1}{\textstyle e^{\textstyle - \nu(T) +p/T} +
1}\quad,\quad
n_{\bar{q}}(\vec{p}\,) = \frac{1}{\textstyle e^{\textstyle \nu(T) +
p/T} + 1},
\end{eqnarray}
where a temperature $T$ is measured in ${\rm MeV}$, $\nu(T) =
\mu(T)/T$, $\mu(T)$ is a chemical potential of the light massless
quarks $q = u,d$, depending on a temperature $T$ [7]. A chemical
potential of light anti--quarks amounts to $- \mu(T)$. A positively
defined $\mu(T)$ provides an abundance of light quarks with respect to
light anti--quarks for a thermalized state [1,4]. A chemical potential
$\mu(T)$ is a phenomenological parameter of the approach which we
would fix below. 

The probability for gluons to have a momentum $\vec{p}$ at a
temperature $T$ is given by the Bose--Einstein distribution function
\begin{eqnarray}\label{label1.2}
n_g(\vec{p}\,) = \frac{1}{\textstyle e^{\textstyle p/T} - 1}.
\end{eqnarray}
Since a strangeness of the colliding heavy--ions amounts to zero,
the densities of strange quarks and anti--quarks should be equal. The
former implies a zero--value of a chemical potential $\mu_s =
\mu_{\bar{s}} = 0$.  In this case the probabilities of strange quarks
and anti--quarks can be given by
\begin{eqnarray}\label{label1.3}
n_s(\vec{p}\,) =
n_{\bar{s}}(\vec{p}\,) = \frac{1}{\textstyle e^{\textstyle
\sqrt{\vec{p}^{\,\,2} + m^2_s}/T} + 1},
\end{eqnarray}
where $m_s =135\,{\rm MeV}$ [8] is the mass of the strange quark and
anti--quark. The value of the current $s$--quark mass $m_s =135\,{\rm
MeV}$ has been successfully applied to the calculation of chiral
corrections to  amplitudes of low--energy interactions, form factors
and mass spectra of low--lying hadrons [9] and charmed heavy--light mesons
[10]. Unlike the massless anti--quarks $\bar{u}$ and $\bar{d}$ for
which the suppression is caused by a chemical potential $\mu(T)$, the
strange quarks and anti--quarks are suppressed by virtue of the
non--zero mass $m_s$.

The multiplicities of the production of $K^+$ and $K^-$ mesons
$N_{K^+}$ and $N_{K^-}$, correspondingly, we describe in the simple
coalescence approach [4,6]. In this case the multiplicities can be
related to the probabilities of the quarks and anti--quarks as
follows:
\begin{eqnarray}\label{label1.4}
N_{K^+}(\vec{q},T) &=& <n_u(\vec{p} - \vec{q}\,)
\,n_{\bar{s}}(\vec{p}\,)>\quad =\nonumber\\
&=&N_C V_K \int\frac{d^3p}{(2\pi)^3}\,\frac{1}{\textstyle
e^{\textstyle - \nu(T) +|\vec{p} - \vec{q}\,|/T} +
1}\frac{1}{\textstyle e^{\textstyle \sqrt{\vec{p}^{\,\,2} + m^2_s}/T}
+ 1},\nonumber\\
N_{K^-}(\vec{q},T) &=& <n_{\bar{u}}(\vec{p} - \vec{q}\,)
\,n_s(\vec{p}\,)>\quad = \nonumber\\
&=&N_C V_K \int\frac{d^3p}{(2\pi)^3}\,
\frac{1}{\textstyle e^{\textstyle \nu(T) + |\vec{p} - \vec{q}\,|/T}
+ 1}\frac{1}{\textstyle e^{\textstyle \sqrt{\vec{p}^{\,\,2} +
m^2_s}/T} + 1},
\end{eqnarray}
where $\vec{q}$ is a 3--momentum of the $K^{\pm}$ mesons, $N_C = 3$
is the number of quark colour degrees of freedom. Then, $V_K$ is a
parameter of a coalescence model [4,6] having dimension of volume and
being to some extent an intrinsic characteristic of spatial
distribution of the $K$ meson. We suggest to determine $V_K$ in terms
of the $K$--meson parameters as follows. Due to the uncertainty
relations the $K$--meson should be localized to the region
proportional to the inverse power of a 3--momentum. For the
thermalized $K$--meson system this should be a thermal momentum. In
the case of the Maxwell--Boltzmann $K$--meson gas the thermal momentum
is proportional to $\sqrt{M_K}$, where $M_K = 500\,{\rm MeV}$ is the
$K$--meson mass.  Another important intrinsic parameter of
pseudoscalar mesons is the leptonic coupling constant $F_P$, $F_K =
160\,{\rm MeV}$ for the $K$ mesons. Thus, from dimensional
consideration we suggest to set $V_K = C/(F_K M_K)^{3/2}$, where $C$
is a dimensionless parameter of the approach equal for all
pseudoscalar mesons. Of course, such a determination of $V_K$ is not
so much rigorous, but it can be useful as a working hypothesis
providing a good agreement with the experimental data.

The multiplicity of the $\pi^+$--meson production we define in an
analogous way:
\begin{eqnarray}\label{label1.5}
N_{\pi^+}(\vec{q},T) &=&<n_u(\vec{p} - \vec{q}\,)
\,n_{\bar{d}}(\vec{p}\,)>\quad =\nonumber\\
&=& N_C V_{\pi}
\int\frac{d^3p}{(2\pi)^3}\,\frac{1}{\textstyle e^{\textstyle -
\nu(T) +|\vec{p} - \vec{q}\,|/T} + 1}\frac{1}{\textstyle
e^{\textstyle  \nu(T) + p/T} + 1},
\end{eqnarray}
where $V_{\pi} = C/(F_{\pi} M_{\pi})^{3/2}$, $F_{\pi} = 131\,{\rm
MeV}$ and $M_{\pi} = 140\,{\rm MeV}$ are the leptonic coupling
constant and the mass of pions.

We should emphasize that other dimensional parameters of $K$ and $\pi$
mesons like charge radii $r_{K^+}$ and $r_{\pi^+}$ cannot be
considered as intrinsic parameters of these mesons, since they are
functions of $F_K$ and $F_{\pi}$. For example, in the Vector Dominance
approach the charge radii can be expressed in terms of the masses of
the $\rho$ and $K^*$ vector mesons $M_{\rho}$ and $M_{K^*}$. According
to KSFR relations imposed by chiral symmetry [9] these masses are
proportional to $F_{\pi}$ and $F_K$, respectively: $M_{\rho} =
g_{\rho}\,F_{\pi} \simeq 790\,{\rm MeV}$ and $M_{K^*}= g_{\rho}F_{K}
\simeq 960\,{\rm MeV}$, where $g_{\rho} \simeq 6$ is the coupling
constant of the $\rho\pi\pi$ interaction. The theoretical values of
the masses of the $\rho$ and $K^*$ mesons predicted by the KSFR
relations agree with experimental values $M_{\rho} = 770\,{\rm MeV}$
and $M_{K^*} = 892\,{\rm MeV}$ within accuracies better than 3$\%$ and
8$\%$, respectively.

The main goal of our paper is to calculating the ratios of the
multiplicities
\begin{eqnarray}\label{label1.6}
R_{K^+K^-}(q, T) &=&
\frac{N_{K^+}(\vec{q},T)}{N_{K^-}(\vec{q},T)},\nonumber\\
R_{K^+\pi^+}(q, T) &=& \frac{N_{K^+}(\vec{q},T)}{N_{\pi^+}(\vec{q},T)}
\end{eqnarray}
with the minimum number of input parameters. Since the main input
parameter of the thermalized approach to the QGP is a chemical
potential $\mu(T)$ of the light quarks and anti--quarks, we would
focus on the possibility to fix it. 

For this aim we suggest to treat a heavy ion as a degenerated Fermi
gas. Converting formally all nucleon degrees of freedom into quark
degrees of freedom we end up with a degenerated Fermi gas of quarks or
differently a degenerated quark--gluon system, where all gluon and
anti--quark degrees of freedom are died out. Heating this quark--gluon
system up to the temperature $T$ and demanding the conservation of the
baryon number, that corresponds to the conservation of the baryon
number density in the fixed spatial volume of the ion, we fix
unambiguously a temperature dependence of a chemical potential of
light quarks and anti--quarks $\mu(T)$ dropping very swiftly at high
temperatures, and the value $\mu(0) = \mu_0 = 250\,{\rm MeV}$.  A
steep falloff of a chemical potential with a temperature implies that
at high temperatures the ratio behaves like $R_{K^+K^-}(q, T) \to
1$. In reality, the limit $R_{K^+K^-}(q, T) \simeq 1$ can be reached
already at $T \ge \mu_0$, where $\mu_0 \simeq 250\,{\rm MeV}$
[1]. However, for intermediate temperatures $T = 160\div 200\,{\rm
MeV}$ as it is estimated in Sect.\,6 the ratio $R_{K^+K^-}(q, T)$
ranges values from the region $2.14 \div 1.48$. This implies that the
experimental analysis of the ratio of the multiplicities of the
$K^{\pm}$--meson production can be a good criterion for the signal of
the QGP phase. Indeed, for the $K^{\pm}$ mesons produced by the
quark--gluon system escaped from the QGP phase the ratio
$N_{K^+}/N_{K^-}$ is expected to be of order of unity.

The paper is organized as follows. In Section\,2 we determine a
chemical potential $\mu(T)$. In Section\,3 we discuss in short a
thermalized QGP.  In Section\,4 we analyse the multiplicities of the
$K^{\pm}$--meson production and give the analytical formulas for the
ratio $R_{K^+K^-}(q,T)$ as a function of 3--momenta of the
$K^{\pm}$ mesons and a temperature $T$. In Section\,5 we analyse the
multiplicities of the $\pi^{\pm}$--meson production and give the
analytical formulas for the ratio $R_{K^+\pi^+}(q,T)$ as a function of
3--momenta of $K^+$ and $\pi^+$ mesons and a temperature $T$. In
Section\,6 we make the numerical analysis of the analytical formulas
derived in Section\,4 and Section\,5.  We show that the ratios depend
smoothly on both 3--momenta of $K^{\pm}$ and $\pi^+$ mesons and the
temperature ranging values from the region $160\,{\rm MeV} \le T \le
200\,{\rm MeV}$. We find that for the temperatures $T = 175\,{\rm
MeV}$ our predictions agree reasonably well with the experimental data
by NA49 and NA44 Collaborations on the central relativistic Pb--Pb
collisions at 158\,GeV per nucleon and the data by NA35 Collaboration
for proton--nucleus and nucleus--nucleus collisions at 200\,GeV per
nucleon.  In the Conclusion we discuss the obtained results.

\section{Chemical potential of light quarks and anti--quarks}
\setcounter{equation}{0}

We suppose that a chemical potential $\mu(T)$, a phenomenological
parameter of the description of the QGP state as a thermalized
quark--gluon system at a temperature $T$, is an intrinsic
characteristic of a thermalized quark--gluon system. Thereby, if the
QGP is an excited state of the QCD vacuum, so a chemical potential
should exist not only for ultra--relativistic heavy--ion
collisions. Quark distribution functions of a thermalized quark--gluon
system at a temperature $T$ should be characterized by a chemical
potential $\mu(T)$ for any external state and any external conditions.
Since any state of a thermalized system is closely related to external
conditions, in order to obtain $\mu(T)$ we need only to specify the
external conditions of a thermalized quark--gluon system the
convenient for the determination of $\mu(T)$.

Let the external conditions of a thermalized quark--gluon system be
caused by the state of the nuclear system, which is a heavy ion with a
baryon number $A$. In the Fermi--gas approximation [11] a heavy ion is
a degenerated gas of nucleons at  $T = 0$ with a baryon
density
\begin{eqnarray}\label{label2.1}
n_B = \frac{A}{\displaystyle \frac{4\pi}{3}r^3_{A}}
 = \frac{3}{4\pi}\,\frac{1}{r^3_N} = 0.14\,{\rm fm}^{-3}
\end{eqnarray}
coinciding with the nuclear matter density $n_{\rm N}$ [11], where
$r_{A} = r_N\,A^{1/3}$ is the radius of a heavy ion, and $r_N =
1.2\,{\rm fm}$ [11,12].

Suppose that all baryon degrees of freedom are converted into quark
degrees of freedom and quarks are massless. In this case we should get
a degenerated Fermi gas of free quarks or more generally a degenerated
free quark--gluon system, where all gluon and anti--quark degrees of
freedom are died out. Heating this quark--gluon system up to a
temperature $T$ we should arrive at a thermalized system of quarks,
anti--quarks and gluons confined in the finite volume of a heavy ion
$(4\pi/3)r^3_N\,A$. 

In the low--temperature limit $T \to 0$ such a
conversion of baryon degrees of freedom into quark ones requiring to
have a system of free quarks can be understood qualitatively, for
example, within a naive non--relativistic quark model, where baryons
are slightly bound three--quark states. These three valence quarks,
the constituent quarks, can be considered as current quark excitations
above a quark condensate produced by a cloud of current $q\bar{q}$
pairs due to spontaneous breaking of chiral symmetry.

In terms of the light quark and anti--quark distribution functions
Eq.(\ref{label1.1}) a baryon density of a thermalized quark--gluon
system at a temperature $T$ is given by [1,4]
\begin{eqnarray}\label{label2.2} 
n_B(T) &=& \frac{1}{3}\times 2\times
2 \times N_C\times [n_q(T) - n_{\bar{q}}(T)] = \nonumber\\
&=&\frac{4}{3}\,N_C\int \frac{d^3p}{(2\pi)^3}\left[\frac{1}{\textstyle
e^{\textstyle - \nu(T) + p/T} +1} - \frac{1}{\textstyle e^{\textstyle
\nu(T) + p/T} +1}\right].  
\end{eqnarray} 
 The factor $(1/3)\times 2 \times 2 \times N_C$ stands for the product
of ({\it baryon charge})$\times$ ({\it number of light quark flavour
degrees of freedom})$\times$({\it number of spin degrees of
freedom})$\times$({\it number of quark colour degrees of
freedom}). The integration over the momentum $\vec{p}$ gives one [1]
\begin{eqnarray}\label{label2.3} 
n_B(T) = \frac{2}{9}\,N_C\,\Bigg[\nu(T)
+ \frac{\nu^3(T)}{\pi^2}\Bigg]\,T^3.  
\end{eqnarray} 
At zero temperature $T = 0$ we get 
\begin{eqnarray}\label{label2.4} 
\mu_0 = \Bigg(\frac{3\pi^2}{2}\Bigg)^{1/3}n^{1/3}_{\rm B} = 250\,{\rm MeV},  
\end{eqnarray} 
where $\mu_0 = \mu(0)$ is a chemical potential at zero temperature,
and $n_{\rm B}(0) = n_{\rm B}$ defined by Eq.(\ref{label2.1}). We have
set $n_{\rm B}(0) = n_{\rm B}$, where $n_{\rm B} =0.14\,{\rm fm}^{-3}$
is given by Eq.(\ref{label2.1}), since in the fixed spatial volume due
to a conservation of a baryon number the baryon density of nucleons
should be equal to the baryon density of quarks. Our result $\mu_0 =
250\,{\rm MeV}$ agrees good with the estimate $\mu_0 \sim 300\,{\rm
MeV}$ [1].

The fluctuations of the quark baryon number $n_B(T)$ caused by the
fluctuations of a temperature $T$, produced in the fixed spatial
volume $(4\pi/3)r^3_N\,A$ of the heavy ion, should lead to the
violation of the baryon number.  As the
baryon number is a good quantum number conserved for strong
interactions, we impose the constraint
\begin{eqnarray}\label{label2.5}
n_B(T) = n_B.
\end{eqnarray}
Form Eq.(\ref{label2.5}) we define the chemical potential $\mu(T)$ as
a function of $T$:
\begin{eqnarray}\label{label2.6}
\frac{\mu(T)}{\mu_0} = \left[\frac{1}{2} + \frac{1}{2}\sqrt{1 +
\frac{4\pi^6}{27}\,\Bigg(\frac{T}{\mu_0}\Bigg)^6}\right]^{1/3} -
\left[-\frac{1}{2} + \frac{1}{2}\sqrt{1 +
\frac{4\pi^6}{27}\,\Bigg(\frac{T}{\mu_0}\Bigg)^6}\right]^{1/3}.
\end{eqnarray}
The dependence of the chemical potential on a temperature $T$ given by
Eq.(\ref{label2.6}) governs the conservation of a baryon number for
the thermalized quark--gluon system confined in the fixed volume $V =
(4\pi/3)\,r^3_N\,A$, when a temperature changes itself. 

In the low--temperature limit $T \to 0$ we get
\begin{eqnarray}\label{label2.7}
\mu(T) = \mu_0\,\Bigg[1 - \frac{\pi^2}{3}\frac{T^2}{\mu^2_0} +
O\Big(T^6\Big)\Bigg].
\end{eqnarray}
The $T$--dependence of a chemical potential given by
Eq.(\ref{label2.7}) differs by a factor $1/4$ from the
low--temperature behaviour of a chemical potential of a thermalized
electron gas [13].

In the high--temperature limit $T \to \infty$ a chemical potential
$\mu(T)$ defined by Eq.(\ref{label2.7}) drops like $T^{-2}$ [1]:
\begin{eqnarray}\label{label2.8}
\mu(T) = \frac{\mu^3_0}{\pi^2}\,\frac{1}{T^2} + O\Big(T^{-7}\Big).
\end{eqnarray}
A chemical potential drops very swiftly when a temperature increases.
Indeed, at $T = 160\,{\rm MeV}$ we obtain $\mu(T) \simeq \mu_0/4$,
while at $T = \mu_0$ a value of a chemical potential makes up
about tenth part of $\mu_0$, i.e. $\mu(T) \simeq \mu_0/10$. This
implies that at very high temperatures the function $\nu(T) = \mu(T)/T$
becomes small  and the contribution of a chemical potential of light
quarks and anti--quarks can be taken into account perturbatively. This
assumes in particular that at temperatures $T \ge \mu_0 = 250\,{\rm MeV}$  the number of light
anti--quarks will not be suppressed by a chemical potential relative to
the number of light quarks.

 Thereby, at very high temperatures the ratio $R_{K^+K^-}(q, T)$
should tend to unity. Hence, the temperature dependence of a chemical
potential given by Eq.(\ref{label2.7}) assumes that hadrons produced
by the quark--gluon system through the QGP phase at temperatures $T >
\mu_0 = 250\,{\rm MeV}$ can be hardly distinguished from the hadrons
procreated by a quark--gluon system escaped from the QGP phase.

However, for intermediate temperatures $T = 160\div 200\,{\rm MeV}$ the
ratio $R_{K^+K^-}(q,T)$ would differ from the unity. Indeed, the rough
estimate gives one 
\begin{eqnarray}\label{label2.9}
R_{K^+K^-}(q,T) \sim {\textstyle e^{\textstyle  2\nu(T)}} =
2.14 \div 1.48 \, >\, 1.
\end{eqnarray}
Hence,  the analysis of the ratio $R_{K^+K^-}(q,T)$ can be still a good 
criterion for the signal of a thermalized QGP phase realized in  
ultra--relativistic heavy--ion collisions.

For the rough estimate of the ratio $R_{K^+\pi^+}(q,T)$ we obtain
\begin{eqnarray}\label{label2.10}
R_{K^+\pi^+}(q,T) \sim \frac{V_K}{V_{\pi}}\,\times\,{\textstyle
e^{\textstyle \nu(T)}} = 0.114 \div 0.059.
\end{eqnarray}
The estimates Eq.(\ref{label2.9}) and Eq.(\ref{label2.10}) are in
qualitative agreement with the experimental data on the central
ultra--relativistic Pb--Pb at 158\,GeV per nucleon collisions by NA49
and NA44 Collaborations and the data on proton--nucleus and
nucleus--nucleus collisions at 200\,GeV per nucleon by NA35
Collaboration: $R^{\exp}_{K^+K^-} = 1.80\pm 0.10$ [14--17] and
$R^{\exp}_{K^+\pi^+} = 0.137\pm 0.008$ [17].

\section{Thermalized quark--gluon plasma}
\setcounter{equation}{0}

In average the QGP phase exists only for the interim of order of
$\tau_{\rm QGP} = (6\div 15)\,{\rm fm/c}$. Therefore,
the thermalization of this system should occur at times $\tau_{\rm
th}$ much less than $\tau_{\rm QGP}$, i.e. $\tau_{\rm QGP} \gg
\tau_{\rm th}$. This can be fulfilled in a very dense matter. Thereby,
in order to be convinced that a thermalized QGP can be realized in
ultra--relativistic heavy--ion collisions we have to calculate $n(T)$,
a particle density of the QGP at a  temperature $T$ containing the
contributions of quarks, anti--quarks and gluons, and to compare the
value of $n(T)$ with the nuclear matter density $n_{\rm N} = 0.14\,{\rm
fm}^{-3}$.

A  particle density of a thermalized QGP  is determined by [1,4]
\begin{eqnarray}\label{label3.1}
n(T) &=& n_g(T) + n_q(T) + n_{\bar{q}}(T) =
2(N^2_C-1)\int\frac{d^3p}{(2\pi)^3}\,\frac{1}{\textstyle e^{\textstyle
p/T} - 1} \nonumber\\
&+& 4\,N_C\int\frac{d^3p}{(2\pi)^3}\,\frac{1}{\textstyle e^{\textstyle
-\nu(T) + p/T} + 1} + 4\,N_C\int\frac{d^3p}{(2\pi)^3}\,\frac{1}{\textstyle
e^{\textstyle \nu(T) + p/T} + 1}.
\end{eqnarray}
Integrating over momenta we obtain
\begin{eqnarray}\label{label3.2}
\hspace{-0.1in}n(T) = T^3\,\frac{4N_C}{\pi^2}\,\Bigg[\Bigg(\frac{N^2_C - 1}{4N_C} + \frac{1}{2}\Bigg)\,\zeta(3) +
\nu^2(T)\,{\ell n}\,2 + \frac{1}{6}\,\nu^3(T) - \int\limits^{\nu(T)}_0
dx\,\frac{(\nu(T) - x)^2}{\textstyle e^{\textstyle x} + 1}\Bigg],
\end{eqnarray}
where $\zeta(3) = 1.202$ is a Riemann zeta--function defined by [18]
\begin{eqnarray}\label{label3.3}
\zeta(s) = \frac{1}{\Gamma(s)}\int\limits^{\infty}_0
dx\,\frac{x^{s-1}}{\textstyle e^{\textstyle x} - 1}
=\frac{1}{\textstyle (1 -
2^{1-s})}\,\frac{1}{\Gamma(s)}\int\limits^{\infty}_0
dx\,\frac{x^{s-1}}{\textstyle e^{\textstyle x} + 1}.
\end{eqnarray}
At $N_C = 3$ the particle density $n(T)$ amounts to
\begin{eqnarray}\label{label3.4}
n(T) = T^3\,\frac{12}{\pi^2}\,\Bigg[\frac{7}{6}\,\zeta(3) + \nu^2(T)\,{\ell
n}\,2 + \frac{1}{6}\,\nu^3(T) - \int\limits^{\nu(T)}_0 dx\,\frac{(\nu(T) -
x)^2}{\textstyle e^{\textstyle x} + 1}\Bigg].
\end{eqnarray}
For temperatures $T \ge 160\,{\rm MeV}$ we can neglect the contribution of
two last terms and with accuracy better than 1$\%$ the particle density is
equal to
\begin{eqnarray}\label{label3.5}
n(T) = T^3\,\frac{12}{\pi^2}\,\Bigg[\frac{7}{6}\,\zeta(3) + \nu^2(T)\,{\ell
n}\,2\Bigg].
\end{eqnarray}
Setting $T \ge 160\,{\rm MeV}$ we get the estimate $n(T) \ge 0.98\,{\rm
fm}^{-3}$. This density is by a factor of order 7  larger compared with the
nuclear matter density $n_{\rm N} = 0.14\,{\rm fm}^{-3}$.

Thus, if we consider quarks, anti--quarks and gluons of a thermalized QGP
like rigid spheres with fixed radii, the average time of collisions
between particles in the QGP is much less than $D/c = (6/\pi n(T)c^3)^{1/3} \simeq  1\,{\rm fm/ c}$, i.e. $\tau_{coll} \ll 1\,{\rm fm/ c}$.  This implies that $\tau_{\rm th}$, the time of a thermalization of the QGP, should be of order $\tau_{\rm th}\le 1\,{\rm fm/ c}$.  This value is of order of magnitude less
compared with the life time of the QGP, i.e. $\tau_{QGP} \ge (6 \div
15)\,\tau_{\rm th}$. These estimates make admissible the application of
a thermalized quark--gluon system to the description of the QGP phase.

For a thermalized QGP the energy density is determined by [1,4]:
\begin{eqnarray}\label{label3.6}
\varepsilon(T) &=& \varepsilon_g(T) + \varepsilon_q(T) +
\varepsilon_{\bar{q}}(T) =
2(N^2_C-1)\int\frac{d^3p}{(2\pi)^3}\,\frac{p}{\textstyle e^{\textstyle
p/T} - 1} \nonumber\\
&+& 4\,N_C\int\frac{d^3p}{(2\pi)^3}\,\frac{p}{\textstyle e^{\textstyle
-\nu(T) + p/T} + 1} + 4\,N_C\int\frac{d^3p}{(2\pi)^3}\,\frac{p}{\textstyle
e^{\textstyle \nu(T) + p/T} + 1}.
\end{eqnarray}
Integrating over momenta we get [1,4]:
\begin{eqnarray}\label{label3.7}
\varepsilon(T) =
T^4\,N_C\,\Bigg[\frac{N^2_C-1}{N_C}\,\frac{\pi^2}{15}+
\frac{7\pi^2}{30} + \nu^2(T) + \frac{1}{2\pi^2}\,\nu^4(T)\Bigg].
\end{eqnarray}
At $N_C = 3$ the energy density of a thermalized QGP amounts to
\begin{eqnarray}\label{label3.8}
\varepsilon(T) = T^4\Bigg[\frac{37 \pi^2}{30} + 3\,\nu^2(T) +
\frac{3}{2\pi^2}\,\nu^4(T)\Bigg].
\end{eqnarray}
Setting $T \ge 160\,{\rm MeV}$ we estimate $\varepsilon(T) \ge
1.08\,{\rm GeV}/{\rm fm}^3$.  Such values of the energy density are
enough for the existence of the QGP [1,4]. Thus, our estimate is on
favour of the existence of a thermalized QGP phase of the quark--gluon
system in ultra--relativistic heavy--ion collisions. Therefore, we can
proceed to the evaluation of the multiplicities of the
$K^{\pm}$-- and $\pi^{\pm}$--meson production caused by a
hadronization of the QGP.

\section{Multiplicities of the $K^{\pm}$--meson production}
\setcounter{equation}{0}

The evaluation of the multiplicities of the $K^{\pm}$--meson
production caused by a hadronization of the QGP we suggest to perform
by making use of Eq.(\ref{label1.4}). Since in our approach to the QGP
quarks, anti--quarks and gluons are described as free Fermi and Bose
gases and the quark--gluon interactions are practically switched off,
we follow a simple coalescence approach to the hadronization
[4,6]. Thus, the multiplicities of the $K^{\pm}$--meson production will
be evaluated through the formulae (see Eq.(\ref{label1.4}))
\begin{eqnarray}\label{label4.1}
N_{K^+}(\vec{q},T) &=& 3\,
V_K\int\frac{d^3p}{(2\pi)^3}\,\frac{1}{\textstyle e^{\textstyle -
\nu(T) +|\vec{p} - \vec{q}\,|/T} + 1}\frac{1}{\textstyle
e^{\textstyle \sqrt{\vec{p}^{\,\,2} + m^2_s}/T} + 1},\nonumber\\
N_{K^-}(\vec{q},T) &=&3\,V_K
\int\frac{d^3p}{(2\pi)^3}\,\frac{1}{\textstyle e^{\textstyle \nu(T) +
|\vec{p} - \vec{q}\,|/T} + 1}\frac{1}{\textstyle e^{\textstyle
\sqrt{\vec{p}^{\,\,2} + m^2_s}/T} + 1},
\end{eqnarray}
where we have set $N_C = 3$.

We show below that the multiplicities given by Eq.(\ref{label4.1}) are
the functions of $\lambda = e^{\textstyle \nu(T)}$, $\lambda_s =
e^{\textstyle m_s/T}$ and $\lambda_K = e^{\textstyle q/T}$:
\begin{eqnarray}\label{label4.2}
N_{K^+}(\vec{q},T) &=& N_{K^+}(\lambda,\lambda_K,\lambda_s),\nonumber\\
N_{K^-}(\vec{q},T) &=&N_{K^+}(\lambda^{-1},\lambda_K,\lambda_s),
\end{eqnarray}
where a 3--momentum $q$ is related to a rapidity $y$, a transversal
momentum $\vec{q}_{\perp}$ and the mass $M_K = 500\,{\rm MeV}$ of the
$K^{\pm}$ mesons as $q = \sqrt{M^2_K{\rm sh}^2y +
\vec{q}^{\,\,2}_{\perp}{\rm ch}^2y}$.

Integrating over the directions of the momentum $\vec{p}$ we obtain
\begin{eqnarray}\label{label4.3}
&&N_{K^+}(\lambda,\lambda_K,\lambda_s) = \frac{3 m^2_s T V_K}{4\pi^2{\ell
n}\,\lambda_K} \int\limits^{\varphi (q)}_0 d\varphi\,\frac{{\rm
sh}\varphi\,{\rm ch}\varphi}{\textstyle 1 + \lambda_s^{\textstyle
{\rm ch}\varphi}}\nonumber\\
&&\times\,\Bigg[\int\limits^{\lambda}_0 dz\,\frac{\textstyle {\ell
n}\Big(1 + z\,\lambda^{-1}_K\,\lambda_s^{\textstyle {\rm
sh}\varphi}\Big)}{z} - \int\limits^{\lambda}_0 dz\,\frac{\textstyle
{\ell n}\Big(1 + z\,\lambda^{-1}_K\,\lambda_s^{\textstyle -{\rm
sh}\varphi}\Big)}{z}\nonumber\\
&&+ ({\ell n}\,\lambda_K - {\ell n}\lambda_s\,{\rm sh}\varphi)\,{\ell
n}\Big( 1 + \lambda\,\lambda^{-1}_K\,\lambda_s^{\textstyle {\rm
sh}\varphi}\Big)\nonumber\\
&&- ({\ell n}\,\lambda_K + {\ell n}\lambda_s\,{\rm sh}\varphi)\,{\ell
n}\Big( 1 + \lambda\,\lambda^{-1}_K\,\lambda_s^{\textstyle - {\rm
sh}\varphi}\Big)\Bigg]\nonumber\\
&&+ \frac{3 m^2_s T V_K}{4\pi^2{\ell
n}\,\lambda_K}\int\limits^{\infty}_{\varphi (q)} d\varphi\,\frac{{\rm
sh}\varphi\,{\rm ch}\varphi}{\textstyle 1 + \lambda_s^{\textstyle
{\rm ch}\varphi}}\nonumber\\
&&\times\,\Bigg[\int\limits^{\lambda}_0 dz\,\frac{\textstyle {\ell
n}\Big(1 + z\,\lambda_K\,\lambda_s^{\textstyle - {\rm
sh}\varphi}\Big)}{z} - \int\limits^{\lambda}_0 dz\,\frac{\textstyle
{\ell n}\Big(1 + z\,\lambda^{-1}_K\,\lambda_s^{\textstyle -{\rm
sh}\varphi}\Big)}{z}\nonumber\\
&&- ({\ell n}\,\lambda_K - {\ell n}\lambda_s\,{\rm sh}\varphi)\,{\ell
n}\Big( 1 + \lambda\,\lambda_K\,\lambda_s^{\textstyle - {\rm
sh}\varphi}\Big)\nonumber\\
&&- ({\ell n}\,\lambda_K + {\ell n}\lambda_s\,{\rm sh}\varphi)\,{\ell
n}\Big( 1 + \lambda\,\lambda^{-1}_K\,\lambda_s^{\textstyle - {\rm
sh}\varphi}\Big)\Bigg],
\end{eqnarray}
where we have also changed a variable $p = m_s\,{\rm sh} \varphi$ and denoted
\begin{eqnarray}\label{label4.4}
\varphi(q) = {\ell n}\Bigg(\frac{q}{m_s} + \sqrt{1 + \frac{q^2}{m^2_s}}\Bigg).
\end{eqnarray}
For the integration over directions $\vec{p}$ we have used the formula
\begin{eqnarray}\label{label4.5}
\int\frac{dx\,x}{\textstyle \lambda^{-1}\,e^{\textstyle x} + 1}= -
x\,{\ell n}\Big(1 + \lambda\,e^{\textstyle - x}\Big) -
\int\limits^{\lambda}_0 dz\,\frac{\textstyle {\ell n}\Big(1 +
z\,e^{\textstyle - x}\Big)}{z}.
\end{eqnarray}
Using Eq.(\ref{label4.2}) we are able to define the ratio of the
multiplicities of the $K^{\pm}$--meson production as follows
\begin{eqnarray}\label{label4.6}
&&R_{K^+K^-}(q,T) =
\frac{N_{K^+}(\lambda,\lambda_K,\lambda_s)}
{N_{K^+}(\lambda^{-1},\lambda_K,\lambda_s)}= \nonumber\\
&&=\Bigg\{\int\limits^{\varphi (q)}_0 d\varphi\,\frac{{\rm sh}\varphi\,{\rm
ch}\varphi}{\textstyle 1 + \lambda_s^{\textstyle {\rm
ch}\varphi}}\,\nonumber\\
&&\times\,\Bigg[\int\limits^{\lambda}_0 dz\,\frac{\textstyle {\ell
n}\Big(1 + z\,\lambda^{-1}_K\,\lambda_s^{\textstyle {\rm
sh}\varphi}\Big)}{z} - \int\limits^{\lambda}_0 dz\,\frac{\textstyle
{\ell n}\Big(1 + z\,\lambda^{-1}_K\,\lambda_s^{\textstyle -{\rm
sh}\varphi}\Big)}{z}\nonumber\\
&&+ ({\ell n}\,\lambda_K - {\ell n}\lambda_s\,{\rm sh}\varphi)\,{\ell
n}\Big( 1 + \lambda\,\lambda^{-1}_K\,\lambda_s^{\textstyle {\rm
sh}\varphi}\Big)\nonumber\\
&&- ({\ell n}\,\lambda_K + {\ell n}\lambda_s\,{\rm sh}\varphi)\,{\ell
n}\Big( 1 + \lambda\,\lambda^{-1}_K\,\lambda_s^{\textstyle - {\rm
sh}\varphi}\Big)\Bigg]\nonumber\\
&&+ \int\limits^{\infty}_{\varphi (q)} d\varphi\,\frac{{\rm
sh}\varphi\,{\rm ch}\varphi}{\textstyle 1 + \lambda_s^{\textstyle
{\rm ch}\varphi}}\,\nonumber\\
&&\times\,\Bigg[\int\limits^{\lambda}_0 dz\,\frac{\textstyle {\ell
n}\Big(1 + z\,\lambda_K\,\lambda_s^{\textstyle - {\rm
sh}\varphi}\Big)}{z} - \int\limits^{\lambda}_0 dz\,\frac{\textstyle
{\ell n}\Big(1 + z\,\lambda^{-1}_K\,\lambda_s^{\textstyle -{\rm
sh}\varphi}\Big)}{z}\nonumber\\
&&- ({\ell n}\,\lambda_K - {\ell n}\lambda_s\,{\rm sh}\varphi)\,{\ell
n}\Big( 1 + \lambda\,\lambda_K\,\lambda_s^{\textstyle - {\rm
sh}\varphi}\Big)\nonumber\\
&&- ({\ell n}\,\lambda_K + {\ell n}\lambda_s\,{\rm sh}\varphi)\,{\ell
n}\Big( 1 + \lambda\,\lambda^{-1}_K\,\lambda_s^{\textstyle - {\rm
sh}\varphi}\Big)\Bigg]\Bigg\}\nonumber\\
&&\times\,\Bigg\{\int\limits^{\varphi (q)}_0 d\varphi\,\frac{{\rm
sh}\varphi\,{\rm ch}\varphi}{\textstyle 1 + \lambda_s^{\textstyle
{\rm ch}\varphi}}\,\nonumber\\
&&\times\,\Bigg[\int\limits^{1/\lambda}_0 dz\,\frac{\textstyle {\ell
n}\Big(1 + z\,\lambda^{-1}_K\,\lambda_s^{\textstyle {\rm
sh}\varphi}\Big)}{z} - \int\limits^{1/\lambda}_0 dz\,\frac{\textstyle
{\ell n}\Big(1 + z\,\lambda^{-1}_K\,\lambda_s^{\textstyle -{\rm
sh}\varphi}\Big)}{z}\nonumber\\
&&+ ({\ell n}\,\lambda_K - {\ell n}\lambda_s\,{\rm sh}\varphi)\,{\ell
n}\Big( 1 + \lambda^{-1}\,\lambda^{-1}_K\,\lambda_s^{\textstyle {\rm
sh}\varphi}\Big)\nonumber\\
&&- ({\ell n}\,\lambda_K + {\ell n}\lambda_s\,{\rm sh}\varphi)\,{\ell
n}\Big( 1 + \lambda^{-1}\,\lambda^{-1}_K\,\lambda_s^{\textstyle - {\rm
sh}\varphi}\Big)\Bigg]\nonumber\\
&&+ \int\limits^{\infty}_{\varphi (q)} d\varphi\,\frac{{\rm
sh}\varphi\,{\rm ch}\varphi}{\textstyle 1 + \lambda_s^{\textstyle
{\rm ch}\varphi}}\,\nonumber\\
&&\times\,\Bigg[\int\limits^{1/\lambda}_0 dz\,\frac{\textstyle {\ell
n}\Big(1 + z\,\lambda_K\,\lambda_s^{\textstyle - {\rm
sh}\varphi}\Big)}{z} - \int\limits^{1/\lambda}_0 dz\,\frac{\textstyle
{\ell n}\Big(1 + z\,\lambda^{-1}_K\,\lambda_s^{\textstyle -{\rm
sh}\varphi}\Big)}{z}\nonumber\\
&&- ({\ell n}\,\lambda_K - {\ell n}\lambda_s\,{\rm sh}\varphi)\,{\ell
n}\Big( 1 + \lambda^{-1}\,\lambda_K\,\lambda_s^{\textstyle - {\rm
sh}\varphi}\Big)\nonumber\\
&&- ({\ell n}\,\lambda_K + {\ell n}\lambda_s\,{\rm sh}\varphi)\,{\ell
n}\Big( 1 + \lambda^{-1}\,\lambda^{-1}_K\,\lambda_s^{\textstyle - {\rm
sh}\varphi}\Big)\Bigg]\Bigg\}^{-1}.
\end{eqnarray}
Thus, due to $\lambda$, $\lambda_K$ and $\lambda_s$ the ratio of the
multiplicities of the $K^{\pm}$--meson production depends on the
3--momenta of the $K^{\pm}$ mesons, i.e. a rapidity $y$ and a
transversal momentum $|\vec{q}_{\perp}|$, and a temperature $T$.

For high momenta $q \to \infty$, i.e. high rapidities $y \to \infty$
or high transversal momenta $|\vec{q}_{\perp}| \to \infty$, the
multiplicities of the $K^{\pm}$--meson production can be substantially
simplified. Indeed, at the limit $q \to \infty$ or the limit
$\lambda_K \to \infty$ the contribution of the integrals over the
region $\varphi(q) \le \varphi < \infty$ can be neglected relative to
the contribution of the integrals over the region $0 \le \varphi \le
\varphi(q)$. Keeping then only the leading terms in large $\lambda_K$
expansion we arrive at the expressions
\begin{eqnarray}\label{label4.7}
N_{K^{\pm}}(\lambda,\lambda_K,\lambda_s) = 
\frac{3 T^3 V_K}{4\pi^2}\,\lambda^{\pm
1}\,\lambda^{-1}_K =
\frac{3 T^2 V_K}{4\pi^2}\,e^{\textstyle \pm \nu(T)}\,{\textstyle
e^{\textstyle - q/T}},
\end{eqnarray}
where the factor $e^{\textstyle - q/T}$ testifies a hadronization 
of the thermalized QGP at a temperature $T$ into a thermalized 
ultra--relativistic $K^{\pm}$--meson gas at a temperature $T$.

Taking then the ratio $R_{K^+K^-}(q,T)$ at the limit $q \to \infty$ we
obtain
\begin{eqnarray}\label{label4.8}
\lim_{q \to \infty} R_{K^+K^-}(q,T) = R_{K^+K^-}(\infty,T) = \lambda^2.
\end{eqnarray}
Hence, in ultra--relativistic heavy--ion collisions going through the
intermediate QGP phase described by the free thermalized quark--gluon
gas at a temperature $T$ the multiplicities of the $K^{\pm}$--meson
production as functions of $\lambda_K$ decrease like
${\ell n}\,\lambda_K/\lambda_K$. In turn the ratio of the
multiplicities $R_{K^+K^-}(q,T)$ does not depend on the momenta of the
$K^{\pm}$--mesons and becomes defined only by a chemical potential of
the light quarks. We show below that the prediction
Eq.(\ref{label4.8}) agrees good with  the experimental data on
central ultra--relativistic Pb--Pb collisions at 158\,GeV per nucleon
in the temperature interval $160\,{\rm MeV} \le T \le 200\,{\rm MeV}$.

The ratio $R_{K^+K^-}(\infty,T)$ given by Eq.(\ref{label4.8}) differs
from the result obtained by Koch, M\"uller and Rafelski (see Eq.(6.29)
of Ref.[4b]) by a factor $\lambda^2_s = \exp(2\mu_s/kT)$, the squared
fugacity of strange quarks, where $\mu_s$ is a chemical potential of
strange quarks. In the case of chemical equilibrium which we follow in
our approach $\mu_s = 0$ and $\lambda_s = \lambda^{-1}_{\bar{s}} = 1$.

In the limit $q \to 0$ the multiplicities of the $K^{\pm}$--meson
production behave like
\begin{eqnarray}\label{label4.9}
N_{K^{\pm}}(\lambda,1,\lambda_s) &=&\frac{3 m^3_s V_K}{2\pi^2}
\int\limits^{\infty}_0 d\varphi\,\frac{{\rm sh}^2\varphi\,{\rm
ch}\varphi}{\textstyle 1 + \lambda_s^{\textstyle {\rm
ch}\varphi}}\,\frac{1}{\textstyle 1 + \lambda^{\mp
1}\,\lambda^{\textstyle {\rm sh}\varphi}_s}.
\end{eqnarray}
It is easy to see that the main contribution to
$N_{K^{\pm}}(\lambda,1,\lambda_s)$ comes from the region $\varphi(q)
\le \varphi < \infty$. Thus, in the limit $q \to 0$ the ratio of the
multiplicities amounts to
\begin{eqnarray}\label{label4.10}
R_{K^+K^-}(0,T) = \Bigg[\int\limits^{\infty}_0 d\varphi\,\frac{{\rm
sh}^2\varphi\,{\rm ch}\varphi}{\textstyle 1 + \lambda_s^{\textstyle
{\rm ch}\varphi}}\,\frac{1}{\textstyle 1 +
\lambda^{-1}\,\lambda^{\textstyle {\rm
sh}\varphi}_s}\Bigg]\,\Bigg[\int\limits^{\infty}_0
d\varphi\,\frac{{\rm sh}^2\varphi\,{\rm ch}\varphi}{\textstyle 1 +
\lambda_s^{\textstyle {\rm ch}\varphi}}\,\frac{1}{\textstyle 1 +
\lambda\,\lambda^{\textstyle {\rm sh}\varphi}_s}\Bigg]^{-1}.
\end{eqnarray}
By varying the momenta of the $K^{\pm}$ mesons over the region $0 \le q <
\infty$ one should get the ratio $R_{K^+K^-}(q,T)$
changing between the values defined by Eq.(\ref{label4.10}) and
Eq.(\ref{label4.8}).

\section{Multiplicity of the $\pi^{\pm}$--meson production}
\setcounter{equation}{0}

In our approach the multiplicities $N_{\pi^+}(\vec{q},T)$ and
$N_{\pi^-}(\vec{q},T)$ of the production of the $\pi^+$ and $\pi^-$
mesons are equal and defined by Eq.(\ref{label1.5}). Integrating over
directions of the momentum $\vec{p}$ we arrive at the expression
depending only on $\lambda$ and $\lambda_{\pi} = e^{\textstyle q/T}$:
\begin{eqnarray}\label{label5.1}
N_{\pi^+}(\lambda, \lambda_{\pi})&=&\frac{3V_{\pi}T^3}{4\pi^2\,{\ell
n}\lambda_{\pi}}\int\limits^{\lambda_{\pi}}_0\frac{dx\,x}{\textstyle
\lambda\,e^{\textstyle x}+ 1}\Bigg[ \int\limits^{\lambda}_0
dz\,\frac{\textstyle {\ell n}\Big(1 +
z\,\lambda^{-1}_{\pi}\,e^{\textstyle x}\Big)}{z} -
\int\limits^{\lambda}_0 dz\,\frac{\textstyle {\ell n}\Big(1 +
z\,\lambda^{-1}_{\pi}\,e^{\textstyle - x}\Big)}{z}\nonumber\\ && +
({\ell n}\,\lambda_{\pi} - x)\,{\ell n}\Big(1 +
\lambda\,\lambda^{-1}_{\pi}\,e^{\textstyle x}\Big)- ({\ell
n}\,\lambda_{\pi} + x)\,{\ell n}\Big(1 +
\lambda\,\lambda^{-1}_{\pi}\,e^{\textstyle -
x}\Big)\Bigg]\nonumber\\ 
&+&\frac{3VT^3}{4\pi^2\,{\ell
n}\lambda_{\pi}}\int\limits^{\infty}_{\lambda_{\pi}}\frac{dx\,x}
{\textstyle
\lambda\,e^{\textstyle x}+ 1}\Bigg[ \int\limits^{\lambda}_0
dz\,\frac{\textstyle {\ell n}\Big(1 +
z\,\lambda_{\pi}\,e^{\textstyle - x}\Big)}{z} -
\int\limits^{\lambda}_0 dz\,\frac{\textstyle {\ell n}\Big(1 +
z\,\lambda^{-1}_{\pi}\,e^{\textstyle - x}\Big)}{z}\nonumber\\ 
&& -
({\ell n}\,\lambda_{\pi} + x)\,{\ell n}\Big(1 +
\lambda\,\lambda^{-1}_{\pi}\,e^{\textstyle - x}\Big)- ({\ell
n}\,\lambda_{\pi} - x)\,{\ell n}\Big(1 +
\lambda\,\lambda^{-1}_{\pi}\,e^{\textstyle - x}\Big)\Bigg].
\end{eqnarray}
The ratio of the production of the $\pi^+$ mesons with respect to the
$K^+$ mesons is determined by the ratio $R_{K^+\pi^+}(q,T)$ which
reads
\begin{eqnarray}\label{label5.2}
&&R_{K^+\pi^+}(q,T) = \frac{N_{K^+}(\lambda,\lambda_K,\lambda_s)}
{N_{\pi^+}(\lambda,\lambda_{\pi})}
=\frac{m^2_s}{T^2}\frac{V_K}{V_{\pi}}\Bigg\{\int\limits^{\varphi (q)}_0
d\varphi\,\frac{{\rm sh}\varphi\,{\rm ch}\varphi}{\textstyle 1 +
\lambda_s^{\textstyle {\rm ch}\varphi}}\,\nonumber\\
&&\times\,\Bigg[\int\limits^{\lambda}_0 dz\,\frac{\textstyle {\ell
n}\Big(1 + z\,\lambda^{-1}_K\,\lambda_s^{\textstyle {\rm
sh}\varphi}\Big)}{z} - \int\limits^{\lambda}_0 dz\,\frac{\textstyle
{\ell n}\Big(1 + z\,\lambda^{-1}_K\,\lambda_s^{\textstyle -{\rm
sh}\varphi}\Big)}{z}\nonumber\\ &&+ ({\ell n}\,\lambda_K - {\ell
n}\lambda_s\,{\rm sh}\varphi)\,{\ell n}\Big( 1 +
\lambda\,\lambda^{-1}_K\,\lambda_s^{\textstyle {\rm
sh}\varphi}\Big)\nonumber\\ &&- ({\ell n}\,\lambda_K + {\ell
n}\lambda_s\,{\rm sh}\varphi)\,{\ell n}\Big( 1 +
\lambda\,\lambda^{-1}_K\,\lambda_s^{\textstyle - {\rm
sh}\varphi}\Big)\Bigg]\nonumber\\ &&+ \int\limits^{\infty}_{\varphi
(q)} d\varphi\,\frac{{\rm sh}\varphi\,{\rm ch}\varphi}{\textstyle 1
+ \lambda_s^{\textstyle {\rm ch}\varphi}}\,\nonumber\\
&&\times\,\Bigg[\int\limits^{\lambda}_0 dz\,\frac{\textstyle {\ell
n}\Big(1 + z\,\lambda_K\,\lambda_s^{\textstyle - {\rm
sh}\varphi}\Big)}{z} - \int\limits^{\lambda}_0 dz\,\frac{\textstyle
{\ell n}\Big(1 + z\,\lambda^{-1}_K\,\lambda_s^{\textstyle -{\rm
sh}\varphi}\Big)}{z}\nonumber\\ &&- ({\ell n}\,\lambda_K - {\ell
n}\lambda_s\,{\rm sh}\varphi)\,{\ell n}\Big( 1 +
\lambda\,\lambda_K\,\lambda_s^{\textstyle - {\rm
sh}\varphi}\Big)\nonumber\\ &&- ({\ell n}\,\lambda_K + {\ell
n}\lambda_s\,{\rm sh}\varphi)\,{\ell n}\Big( 1 +
\lambda\,\lambda^{-1}_K\,\lambda_s^{\textstyle - {\rm
sh}\varphi}\Big)\Bigg]\Bigg\}\nonumber\\
&&\times\,\Bigg\{\int\limits^{{\ell
n}\,\lambda_{\pi}}_0\frac{dx\,x}{\textstyle
\lambda\,e^{\textstyle x}+ 1}\Bigg[ \int\limits^{\lambda}_0
dz\,\frac{\textstyle {\ell n}\Big(1 +
z\,\lambda^{-1}_{\pi}\,e^{\textstyle x}\Big)}{z} -
\int\limits^{\lambda}_0 dz\,\frac{\textstyle {\ell n}\Big(1 +
z\,\lambda^{-1}_{\pi}\,e^{\textstyle - x}\Big)}{z}\nonumber\\ && +
({\ell n}\,\lambda_{\pi} - x)\,{\ell n}\Big(1 +
\lambda\,\lambda^{-1}_{\pi}\,e^{\textstyle x}\Big)- ({\ell
n}\,\lambda_{\pi} + x)\,{\ell n}\Big(1 +
\lambda\,\lambda^{-1}_{\pi}\,e^{\textstyle -
x}\Big)\Bigg]\nonumber\\ && + \int\limits^{\infty}_{{\ell
n}\,\lambda_{\pi}}\frac{dx\,x} {\textstyle
\lambda\,e^{\textstyle x}+ 1}\Bigg[ \int\limits^{\lambda}_0
dz\,\frac{\textstyle {\ell n}\Big(1 +
z\,\lambda_{\pi}\,e^{\textstyle - x}\Big)}{z} -
\int\limits^{\lambda}_0 dz\,\frac{\textstyle {\ell n}\Big(1 +
z\,\lambda^{-1}_{\pi}\,e^{\textstyle - x}\Big)}{z}\nonumber\\ &&-
({\ell n}\,\lambda_{\pi} - x)\,{\ell n}\Big(1 +
\lambda\,\lambda_{\pi}\,e^{\textstyle - x}\Big) - ({\ell
n}\,\lambda_{\pi} + x)\,{\ell n}\Big(1 +
\lambda\,\lambda^{-1}_{\pi}\,e^{\textstyle -
x}\Big)\Bigg]\Bigg\}^{-1}.
\end{eqnarray}
For high momenta $q \to \infty$ the multiplicity of the
$\pi^{\pm}$--meson production behave like the multiplicity of the
$K^{\pm}$--meson production
\begin{eqnarray}\label{label5.3}
N_{\pi^{\pm}}(\lambda,\lambda_{\pi})= \frac{3 T^3
V_{\pi}}{4\pi^2}\,\lambda^{-1}_{\pi} = \frac{3 T^3
V_{\pi}}{4\pi^2}\,{\textstyle e^{\textstyle - q/T}},
\end{eqnarray}
where the factor $e^{\textstyle - q/T}$ testifies a hadronization of
the thermalized QGP at a temperature $T$ into a thermalized
ultra--relativistic $\pi^{\pm}$--meson gas at a temperature $T$.

Taking into account Eq.(\ref{label4.7}) we obtain the ratio
$R_{K^+\pi^+}(q, T)$ at the limit $q \to \infty$:
\begin{eqnarray}\label{label5.4}
\lim_{q \to \infty}R_{K^+\pi^+}(q, T) = R_{K^+\pi^+}(\infty, T) =
 \frac{V_K}{V_{\pi}}\,e^{\textstyle \nu(T)}.
\end{eqnarray}
Below we show that this relation agrees good with the experimental
data on nucleus--nucleus ultra--relativistic collisions at 200\,GeV
per nucleon (NA35 Collaboration).

It is remarkable that in the ratio a parameter $C$ is canceled and the
ratio $R_{K^+\pi^+}(q, T)$ is determined by good established
parameters of the $K^+$ and $\pi^+$ mesons such as the masses and the
leptonic coupling constants.

\section{Numerical analysis}
\setcounter{equation}{0}

The numerical analysis of the ratio $R_{K^+K^-}(q,T)$ given by
Eq(\ref{label4.6}) displays that it slightly varies around the value
$R_{K^+K^-}(\infty,T) = \lambda^2$, when 3--momenta of the
$K^{\pm}$ mesons take  values from the region $0 \le q < 10^3\,{\rm
GeV}$, and it satisfies the relation Eq.(\ref{label4.8}) at $q \ge
10^3\,{\rm GeV}$.

The ratio Eq.(\ref{label4.6}) depends
also smoothly on a temperature ranging over the region $160\,{\rm MeV}
\le T < 200\,{\rm MeV}$. The numerical data read
\begin{eqnarray}\label{label6.1}
R_{K^+K^-}(q,T=160) &=& 2.14^{+0.13}_{-0.30},\nonumber\\
R_{K^+K^-}(q,T=175) &=& 1.80^{+0.04}_{-0.18},\nonumber\\
R_{K^+K^-}(q,T=190) &=& 1.58^{+0.02}_{-0.13},
\end{eqnarray}
where the upper and the lower values correspond to the maximum and the
minimum of the ratio, respectively.

When matching the theoretical values Eq.(\ref{label6.1}) with the
experimental data on central ultra--relativistic PB--Pb collisions at
158/,GeV per nucleon [14--17]:
\begin{eqnarray}\label{label6.2}
R^{\exp}_{K^+K^-} = 1.80\pm 0.10,
\end{eqnarray}
one can see that our approach describes good the experimental data on
the production of the $K^{\pm}$--mesons at the temperature $T =
175\,{\rm MeV}$.

Since in the case of a hadronization of a quark--gluon system escaped
from the QGP phase we have obtained the ratio $R_{K^+K^-}$ independent
on the momentum of the $K^{\pm}$--mesons and equal to $R_{K^+K^-} =
1.10\pm 0.01$, the numerical estimates Eq.(\ref{label6.1}) evidently
testify that the intermediate state in ultra--relativistic heavy--ion
collisions should run via the QGP phase with a handsome probability.

For the entire region of the 3--momenta $0 \le q < \infty$ the ratio
$R_{K^+\pi^+}(q,T)$ of the multiplicities of the $K^+$ and $\pi^+$
meson production varies very smoothly and is given by
\begin{eqnarray}\label{label6.3}
R_{K^+\pi^+}(q,T=160) &=& 0.144 \pm 0.017,\nonumber\\
R_{K^+\pi^+}(q,T=175) &=& 0.134\pm 0.014,\nonumber\\
R_{K^+\pi^+}(q,T=190) &=& 0.128\pm 0.011,
\end{eqnarray}
where $\pm \Delta$ gives the minimum and the maximum values of the
ratio. When matching the theoretical values of the ratio
Eq.(\ref{label6.3}) with the experimental data, given by NA35
Collaboration on proton--nucleus and nucleus--nucleus collisions at
200\,GeV per nucleon [17]
\begin{eqnarray}\label{label6.4}
R^{\exp}_{K^+\pi^+} = 0.137\pm 0.008,
\end{eqnarray}
one can see  that the best agreement we get again for $T = 175\,{\rm MeV}$.

\section{Conclusion}
\setcounter{equation}{0} 

The main point of our approach to the description of the QGP phase as
a thermalized quark--gluon system is in the determination of a
chemical potential of light quarks $\mu(T)$. We have defined $\mu(T)$
as a function of a temperature $T$ and calculated its value at $T = 0$
in terms of the baryon density of nucleons $n_{\rm B}$, which
coincides with the nuclear matter density $n_{\rm B} = n_{\rm N} =
0.14\,{\rm fm}^{-3}$. Our result for the chemical potential at zero
temperature $\mu_0 = 250\,{\rm MeV}$ agrees good with the estimate
$\mu_0 \sim 300\,{\rm MeV}$ [1].

For the evaluation of multiplicities of the $K^{\pm}$--meson
production we have followed a simple coalescence model defining
multiplicities in terms of geometrical probabilities for pairs
$u\bar{s}\,$(or $\bar{u}s$) to be at a spatial volume $V_K$, where
probabilities quarks and anti--quarks are given by Fermi distribution
functions at a temperature $T$. We have shown that in the
ultra--relativistic limit these multiplicities acquire the shape of
the Maxwell--Boltzmann distribution functions testifying the existence
of a thermalized $K^{\pm}$--meson gas at temperature $T$ in the
hadronic phase of the thermalized QGP.

The ratio of these multiplicities $R_{K^+K^-}(q,T)$ has turned out to
be a smooth function on 3--momenta $q$ of the $K^{\pm}$ mesons in
the entire region $0 \le q \le \infty$ varying slightly around the
values $R_{K^+K^-}(\infty,T) = e^{\textstyle 2 \mu(T)/T}$. At
$T=175\,{\rm MeV}$ we have found a value $R_{K^+K^-}(q,T=175)=
1.80^{+0.04}_{-0.18}$ agreeing good with the experimental data on
central ultra--relativistic Pb--Pb collisions at 158 GeV per nucleon
(NA49 and NA44 Collaborations) and ultra--relativistic nucleus--nucleus
collisions at 200 GeV per nucleon (NA35 Collaboration) [14--17]:
$R^{\exp}_{K^+K^-} = 1.80\pm 0.10$. Therewith, the thermodynamical
parameters of our fit of experimental data $T = 175\,{\rm MeV}$ and
$\mu(T=175) = 51\,{\rm MeV}$ are in qualitative agreement with the
experimental ones [15]: $T \sim 170\,{\rm MeV}$ and $\mu \sim
85\,{\rm MeV}$.

The ratio of the multiplicities of the $K^{\pm}$--meson production in
the thermalized QGP has been calculated as a function of a chemical
potential of light quarks and a temperature by Koch, M\"uller and
Rafelski [4b]. When matching our result given for the ratio
$R_{K^+K^-}(q,T)$ by Eqs.(\ref{label4.6}) and (\ref{label6.1}) with
the result obtained by Koch, M\"uller and Rafelski (see Eq.(6.29) and
Fig.\,6.7 of Ref.[4b]) we argue that in the QGP phase of QCD (i)
strange quark and anti--quarks are in the equilibrium state that
provides a vanishing value of their chemical potential, (ii) the ratio
of multiplicities is a smooth function of 3--momenta of $K^{\pm}$
mesons, (iii) the most important values of a temperature are higher
than $T = 160\,{\rm MeV}$ and (iv) a chemical potential of light
quarks is a swiftly dropping with $T$ function. Then, we would like to
emphasize that unlike the approach given by Koch, M\"uller and
Rafelski Ref.[4b] in our model there are no free parameters.  This
makes the obtained results much more sensitive when compared with the
experimental data, and the agreement with the experimental data, if
reached, should signify a correct mechanism of the QGP. Indeed, the
ratio $R_{K^+K^-}(q,T)$ depending only on a temperature can be
unambiguously compared with the experimental data by varying only a
temperature $T$ in reasonable limits.

The ratio $R_{K^+\pi^+} (q, T)$ of the multiplicities of the $K^+$ and
$\pi^+$--meson production has been found as a smooth function of the
3--momenta $0 < q < \infty$.  The absolute value of the ratio
$R_{K^+\pi^+} (q, T)$ depends on the parameter of the model
$V_K/V_{\pi}$. The ratio $V_K/V_{\pi}$ is fixed in our approach in
terms of good established parameters of the $K^+$ and $\pi^+$ mesons:
$V_K/V_{\pi} = (F_{\pi}M_{\pi}/F_KM_K)^{3/2} = 0.109$. This gives the
ratio $R_{K^+\pi^+} (q, T)$ which agrees good with experimental data
given by NA35 Collaboration [17]: $R^{\exp}_{K^+\pi^+} = 0.137\pm
0.008$. The best agreement we get at $T = 175\,{\rm MeV}$:
$R_{K^+\pi^+}(q,T = 175) = 0.134\pm 0.014$. At $T= 175\,{\rm MeV}$ the
ratio $R_{K^+\pi^+} (q, T)$ changes from a minimal value $R_{K^+\pi^+}
(q, T) = 0.120$ to a maximal one $R_{K^+\pi^+} (q, T) = 0.148$ with a
3--momentum ranging over the region $q \in [0,\infty)$.

First calculation of the ratio of the multiplicities of the $K^+$-- and
$\pi^+$--meson production in the thermalized QGP has been performed by
Glendenning and Rafelski [19]. Unlike our approach in Ref.[19] for the
description of the multiplicity of the production of $\pi^+$ mesons
there has been used the distribution function corresponding to the
consideration of the produced $\pi^+$ mesons as a thermalized Bose
gas. To some extent this leads to a loss of the quark origin of
$\pi^+$ mesons which is retained in our approach Eq.(\ref{label1.5}).
When comparing the numerical value of the ratio $R_{K^+\pi^+} (q, T)$
obtained in our model with that calculated by Glendenning and Rafelski
$K^+/\pi^+\approx 0.3$ (see Fig.\,2  of Ref.[19]) for
$T = 160\div 180\,{\rm MeV}$ we should testify that our result agrees
better with the contemporary experimental data.

Such an agreement obtained for $R_{K^+K^-} (q, T)$ and $R_{K^+\pi^+}
(q, T)$ seems to be rather interesting, since in our approach there is
no free parameters save a temperature $T$. Indeed, a chemical
potential at zero temperature $\mu_0 = 250\,{\rm MeV}$ is determined
through the nuclear matter density $n_{\rm N} = 0.14\,{\rm
fm}^{-3}$. Then, the mass of a current $s$--quark $m_s = 135\,{\rm
MeV}$ is quoted by QCD at the normalization scale $1\,{\rm GeV}$ [8],
and it has been successfully applied to numerous calculations of fine
chiral structure of both light and heavy--light mesons [9,10]. This
implies that our definition of a chemical potential $\mu(T)$ carried
out through the requirement of a conservation of a heavy ion baryon
number at any temperature $T$ describes good a thermodynamical
properties of the QGP in the form of a thermalized free Fermi--Bose
gas of quarks, anti--quarks and gluons. Then, an arbitrary parameter
$C$ entering to the definition of the parameters $V_K$ and $V_{\pi}$,
the coalescence model parameters, has been canceled in the ratio
$V_K/V_{\pi}$. As a result the ratio $R_{K^+\pi^+} (q, T)$ has been
defined by good established parameters of $K$ and $\pi$ mesons: the
masses $M_{\pi} = 140\,{\rm MeV}$ and $M_K = 500\,{\rm MeV}$, and the
leptonic coupling constants $F_{\pi} = 131\,{\rm MeV}$ and $F_K =
160\,{\rm MeV}$.

In the case of a hadronization of a quark--gluon system escaped from
the QGP phase we have obtained the ratio $R_{K^+K^-}$ independent on
the momentum of the $K^{\pm}$--mesons and equal to $R_{K^+K^-} =
1.10\pm 0.01$. Therefore, the numerical estimates of the ratio
$R_{K^+K^-}(q, T)$ carried out in out approach to the description of
the QGP evidently testify that the intermediate state in
ultra--relativistic heavy--ion collisions should run via the QGP phase
with a handsome probability.

Our success in describing of the ratios of the $K^{\pm}$--
and $\pi^{\pm}$--meson productions in ultra--relativistic heavy--ion
collisions with a minimal number of free parameters should be
supported by the description of the ratios of the strange baryon and
anti--baryon production [20]:
$R_{\bar{\Lambda}/\bar{\Sigma}\,\Lambda/\Sigma}= 0.133\pm 0.007$,
$R_{\bar{\Xi}\,\Xi}=0.249\pm 0.019$ and $R_{\bar{\Omega}\,\Omega} =
0.383\pm 0.081$. The analysis of these experimental data in our
approach are planned in our forthcoming publications.

When comparing our approach with others we would like to refer to the
paper by Bir$\acute{\rm o}$ and Ziman$\acute{\rm y}$i [4a]. Indeed,
our constraint $n_{\rm B}(T) = n_{\rm B}$ given by
Eq.(\ref{label2.5}), allowing to determine a temperature dependence of
a chemical potential, corresponds to some extent to the relation
Eq.(2.18) of Ref.[4a]. This relation can be obtained from
Eq.(\ref{label2.5}) by multiplying the r.h.s. of Eq.(\ref{label2.5})
by the factor $J\gamma_0$, where $\gamma_0$ is a Lorentz factor
related to the laboratory energy of colliding heavy ions and $J$ is a
parameter of the approach [4a]. This gives $n_{\rm B}(T) = n_{\rm
B}\,J\,\gamma_0$. According to Ref.[4a] the factor $J\,\gamma_0$ is
always greater than unity $J\,\gamma_0 >1$, whereas in our approach
$J\gamma_0 = 1$. Unlike our approach, where all parameters are fixed,
the parameter $J$ is a free parameter of the model [4a].

\section{Acknowledgement}

We are grateful to Prof. Bir$\acute{\rm o}$ and Prof. Pi{\v s}$\acute{\rm u}$t for helpful discussions.

\end{document}